\def \SAIT #1 #2 {{\em Mem.\ Soc.\ Astron.\ It.\/} {\bf #1}, #2}
\def \MESS #1 #2 {{\em The Messenger\/} {\bf #1}, #2}
\def \ASTRNACH #1 #2 {{\em Astron. Nach.\/} {\bf #1}, #2}
\def \AAP #1 #2 {{\em Astron. Astrophys.\/} {\bf #1}, #2}
\def \AAL #1 #2 {{\em Astron. Astrophys. Lett.\/} {\bf #1}, L#2}
\def \AAR #1 #2 {{\em Astron. Astrophys. Rev.\/} {\bf #1}, #2}
\def \AAS #1 #2 {{\em Astron. Astrophys. Suppl. Ser.\/} {\bf #1}, #2}
\def \AJ #1 #2 {{\em Astron. J.\/} {\bf #1}, #2}
\def \ANNREV #1 #2 {{\em Ann. Rev. Astron. Astrophys.\/} {\bf #1}, #2}
\def \APJ #1 #2 {{\em Astrophys. J.\/} {\bf #1}, #2}
\def \APJL #1 #2 {{\em Astrophys. J. Lett.\/} {\bf #1}, L#2}
\def \APJS #1 #2 {{\em Astrophys. J. Suppl.\/} {\bf #1}, #2}
\def \APSS #1 #2 {{\em Astrophys. Space Sci.\/} {\bf #1}, #2}
\def \ASR #1 #2 {{\em Adv. Space Res.\/} {\bf #1}, #2}
\def \BAIC #1 #2 {{\em Bull. Astron. Inst. Czechosl.\/} {\bf #1}, #2}
\def \JSQRT #1 #2 {{\em J. Quant. Spectrosc. Radiat. Transfer\/} {\bf #1}, #2}
\def \MN #1 #2 {{\em Mon. Not. R. Astr. Soc.\/} {\bf #1}, #2}
\def \MEM #1 #2 {{\em Mem. R. Astr. Soc.\/} {\bf #1}, #2}
\def \PLR #1 #2 {{\em Phys. Lett. Rev.\/} {\bf #1}, #2}
\def \PASJ #1 #2 {{\em Publ. Astron. Soc. Japan\/} {\bf #1}, #2}
\def \PASP #1 #2 {{\em Publ. Astr. Soc. Pacific\/} {\bf #1}, #2}
\def \NAT #1 #2 {{\em Nature\/} {\bf #1}, #2}
\def \etal {{\em et al.\/} }

\documentstyle{memsait}
\input epsf.sty
\begin{opening}
\title{MAPPING THE ARTIFICIAL SKY BRIGHTNESS IN EUROPE FROM DMSP SATELLITE MEASUREMENTS: THE SITUATION OF THE NIGHT SKY IN ITALY IN THE LAST QUARTER OF CENTURY.} 
\author{PIERANTONIO CINZANO$^{1}$, FABIO FALCHI$^{1}$, CHRISTOPHER D. ELVIDGE$^{2}$, KIMBERLY E. BAUGH$^{2}$}
\institute{$^{1}$ Dipartimento di Astronomia, Universit\`a di Padova,
vicolo dell'Osservatorio 5,  I-35122 Padova, Italy, 
email: cinzano@pd.astro.it, falchifa@tin.it\\
$^{2}$ Solar-Terrestrial Physics Division, NOAA National Geophysical Data Center, 3100 Marine Street, Boulder CO 80303
}
\date{} 
\end{opening}

\begin{document}

\oddpagefooter{}{}{} 
\evenpagefooter{}{}{} 
\ 
\bigskip

\begin{abstract}
We present a project to map the artificial sky brightness in Europe
in the main astronomical photometrical bands with a resolution better than 3 km. The aim is to understand the state of night sky pollution in Europe, to quantify the present situation and to allow future monitoring of trends. The artificial sky brightness in each site at a given position on the sky is obtained by the integration of the contributions produced by every surface area in the surroundings of the site. Each contribution is computed taking in account the propagation in the atmosphere of the upward light flux emitted by the area and measured from DMSP satellites. The project is a long term study in which we plan to take in account successively of many different details in order to improve the maps. 

We present, as a preliminary result, a map of the V-band artificial sky brightness in Italy in 1998 and we compare it with the map obtained 27 years earlier by Bertiau, Treanor and De Graeve. Predictions for the artificial sky brightness within the next 27 years are also shown. 
\end{abstract}


\section{Introduction}

 Mappings of sky brightness for extended areas were performed by Walker (1970, 1973), Albers (1998) in USA, Bertiau, de Graeve and Treanor (1973; Treanor 1974) in Italy  and Berry (1976) in Canada with some simple modelling. These authors used population data of cities to estimate their upward light emission and a variety of propagation laws for light pollution in order to compute the sky brightness. Recently DMSP satellite images allowed direct information on the upward light emission from almost all countries around the World (Sullivan 1989, 1991; Elvidge \etal 1997a, 1997b, 1998) and were used to study the increase of this flux with time (Isobe 1993; Isobe \& Hamamura 1998).
 
 In this paper we present a project to obtain detailed maps of artificial sky brightness in Europe in astronomical photometrical bands with a resolution better than 3 km. In order to bypass errors  arising when using population data to estimate upward flux, we construct the maps measuring directly the upward flux in DSMP satellite night-time images and convolving it with a light pollution propagation function. 
 
DMSP are satellites of the Defense Meteorological Satellite Program (DMSP) of the National Oceanic and Atmospheric Administration (NOAA) in a low altitude (830 km) sun-synchronous polar orbit with an orbital period of 101 minutes. Visible and infrared imagery from DMSP Operational Linescan System (OLS) instruments monitor twice a day, one in daytime and one in nightime, the distribution of clouds all over the world. At night the instrument for visible imagery is a Photo Multiplier Tube (PMT) sensitive to radiation from 410 nm to 990 nm (470-900 FWHM) with the highest sensitivity at 550-650 nm, where the most used lamps for external night-time lighting have the strongest emission: Mercury Vapour (545 nm and 575 nm), High Pressure Sodium (from 540 nm to 630 nm), Low Pressure Sodium (589 nm). The IR detector is sensitive to radiation from 10,0 $\mu m$ to 13,4 $\mu m$ (10.3-12.9 FWHM).  Every fraction of a second each satellite scans a narrow swath extending 3000 km in east-west direction. Data received by NOAA National Geophysical Data Center have a nominal spatial resolution of 2.8km obtained by on-board averaging of five by five blocks of finer data with spatial resolution of 0.56km.
 
 \section{Description of the method}
 
Main steps of our method are:
 \begin{enumerate}
  
 \item {\it Construction of an expanded dynamics composite image.} 
We search for images cloudfree as guaranteed by inspection of the IR images taken at the same time. In fact the presence of clouds over some cities constitute a possible source of error: these clouds could hide or dim the light received by the satellite.  Due to the limited dynamic range of the satellite detectors, the automatic gain normally saturates the most lit pixel of the largest cities. Sensitivity reaches $10^{-5}$ W $m^{-2} ~sr^{-1} ~\mu m^{-1}$ (Elvidge et al. 1997). Few images sometime are taken with lower gain (50-24 db) and they have only a few saturated pixels. We construct a composite image replacing the saturated pixels in the higher gain images, useful to measure accurately low population sources, with the measurements coming from the low gain images, adequately rescaled. The pixel values are currently relative values rather than absolute values because instrumental gain levels are adjusted to have a constant cloud reference brightness in different lighting conditions related to the solar and lunar illumination at the time. Elvidge et al(1998) obtained a calibrated composite image of the entire word in 1998 from images specially taken without gain. A possible source of errors on satellite measurements is that each pixel of the 2.8km resolution images is the sum of smaller detector pixels and we haven't any way to check if some of them were saturated. This uncertainty will be solved only when high resolution images will be at our disposal. 
 
 \item {\it Estimate of the upward light flux.} We analyze the composite image measuring pixel counts. Under the hypotesis of uniformity of the shape of the average upward emission function we obtain the relative upward light flux of the area covered by each pixel. 
 
 \item {\it Propagation of the upward light flux in the atmosphere.} The scattering from atmospheric particles and molecules spreads the light emitted upward by the cities. If $f((x,y),(x',y'))$ is a propagation law for light pollution giving the artificial sky brightness produced at a given position of the sky in a site in $(x',y')$ by an infinitesimal area  $dS=dxdy$ in $(x,y)$ with unitary upward emission per unit area, the total artificial sky brightness $b$ at that position in the site is given by:
\begin{equation}
\label{int1}
b(x',y')=\int\int e(x,y) f((x,y),(x',y'))~dx ~dy
\end{equation}
This expression is the convolution of the upward emission per unit area $e(x,y)$ with the propagation function $f((x,y),(x',y'))$. 
In pratice, we divide the surface of Europe in pixels with the same positions and dimensions as in the satellite image. We assume each area of the country defined by a pixel be a source of light pollution with an upward emission $e_{x,y}$ proportional to the measured pixel counts. In this case the sky brightness at the center of each pixel given by the expression (\ref{int1}) became:
\begin{equation}
b_{i,j}=\sum_{h}\sum_{k}  e_{h,k} f((x_{i},y_{j}),(x_{h},y_{k}))
\end{equation} 
 The propagation function $f((x_{i},y_{j}),(x_{h},y_{k}))$ for each couple of points $(x_{i},y_{j})$ and $(x_{h},y_{k})$ (the positions of the observing site and the polluting area) is obtained with detailed models based on the modelling technique introduced and developed by Garstang (1986, 1987, 1988, 1989a, 1989b, 1989c, 1991a, 1991b, 1991c, 1992, 1993, 1999) and also applied by Cinzano (1999a, 1999b, 1999c). 
 For each infinitesimal volume of atmosphere along the line-of-sight, the direct illuminance produced by each source and the illuminance due at light scattered once from molecules and aerosols are computed, this last estimed with the approach of Treanor (1973) as extended by Garstang (1984, 1986). The total flux that molecules and aerosols in the infinitesimal volume scatter toward the observer is computed from the illuminance, and, with an integration, the artificial sky brightness of the sky in that direction is obtained. Extinction along light paths is taken in account. The model assumes Rayleigh scattering by molecules and Mie scattering by aerosols. It is possible take in account of the altitudes of sites and sources, the Earth Curvature effects, the scattering functions in the choosen photometrical band, the shape of average city upward emission functions, and their geographical gradients.  These models allow to associate the predictions to well-defined parameters related to the aerosol content, so the atmospheric conditions at which predictions refer can be well known.  
 
 \item {\it Calibration of the results for a choosen aerosol content.} Except that in the case of Elvidge \etal (1998), usually satellite images  are not calibrated so upward flux measurements and artificial sky brightness maps are only relative. We calibrate the maps on the basis of (i) analysis of existing radiance calibrated satellite images (Cinzano, Falchi, Elvidge, Serke, in prep.) or/and (ii) accurate measurements of sky brightness/luminance together with estinction from the earth-surface (e.g. Falchi, Cinzano 1999).
 \end{enumerate}

 \section{Subsequent upgrades}
  
We plan to take in account subsequently of:
 \begin{enumerate}
 \item Altitude of each area. At first we plan to compute the maps for sea level.
 \item Photometrical Bands. We plan to start with V and B photometrical bands and to extend to R and U bands later. Astronomical brightness in V mag/arcsec$^{2}$d can be transformed in luminance in cd/m$^{2}$.
\item Space resolution. We will start from $\sim$2.8 km resolution but we hope to go down to $\sim$0.5 km as soon high resolution images will be obtained.
\item Curvature of the earth. This might produce an error for isolated areas but in strongly urbanized areas it is negligible. The effect of earth curvature is about 2 percent at 50 km (Garstang 1989). We will extend our calculations later.
 \item Geographical gradients of the atmospheric aerosol content. The same atmospheric model as Garstang (1996, 1991) (also used by Cinzano 1999a; 1999b; 1999c) will be assumed at first, with the density of molecules and aerosols decreasing exponentially with the height. We plan to use better atmospheric models as soon as available.
 \item Geographical gradients of the average aerosol angular scattering function. The average scattering function of aerosols that will be adopted at first, is the representation of Garstang (1991) of the function measured by McClatchey \etal (1978). 
 \item Geographical gradients of the average upward flux emission function of cities. In order to became simple, we  will assume at first that the lighting habits are similar in all the cities.
 Some studies has been undertaken to determinate the better average function from satellite data (Falchi and Cinzano, in prep.) and from Earth-based observations (Cinzano, in prep.).
 \end{enumerate}
 
To account for presence of sporadic denser aerosol layers at various heights or at ground level as Garstang (1991b) is beyond the scope of this work. We will also neglect the presence of mountains which might shield the light emitted from the sources to a fraction of the atmospheric particles along the line-of-sight of the observer. Given the vertical extent of the atmosphere in respect to the highness of the mountains, the shielding is not negligible only when the source is very near the mountain and both are quite far from the site (Garstang 1989, see also Cinzano 1999a). We also neglected the effects of the Ozone layer and the presence of volcanic dust studied by Garstang (1991b, 1991c).

\section{Sky brightness in Italy in 1971 and in 1998.}

In order to test our method we obtained a first map of the V-band zenith artificial sky brightness in Italy (Falchi and Cinzano 1999; Falchi 1999) applying the Treanor Law, a very simple light pollution propagation function obtainable simplifying the Garstang integral under the hypotesis of Treanor (1973): 
\begin{figure}
\epsfysize=19cm 
\hspace{1.5cm}\epsfbox{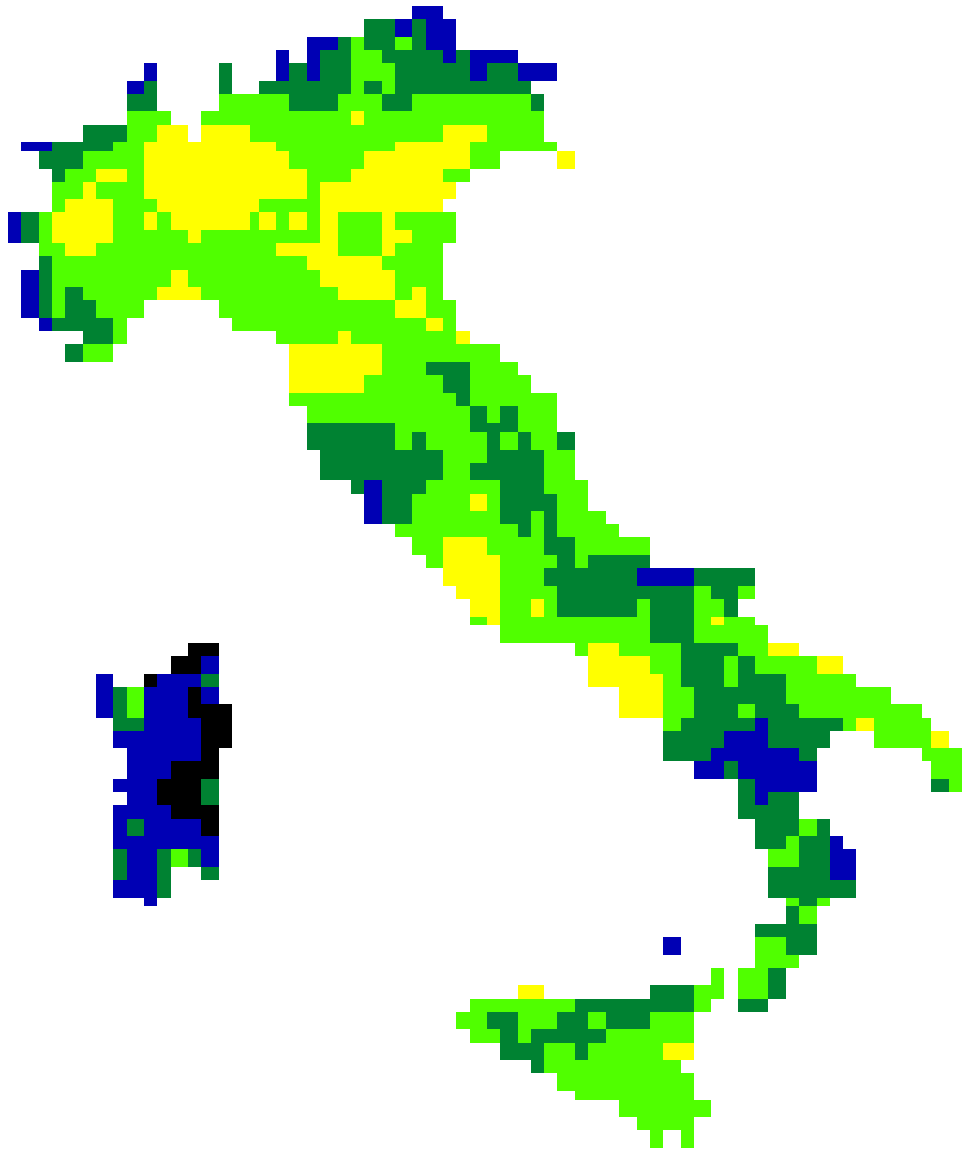} 
\caption[h]{Map of the artificial sky brightness in Italy in 1971 from data from Bertiau {\it et al.} (1973). Levels correspond respectively to $<$0.05, 0.05-0.15, 0.15-0.35, 0.35-1.1, $<$1.1 times the natural sky brightness. }
\label{fig1}
\end{figure}

\begin{figure}
\epsfysize=19cm 
\hspace{1.5cm}\epsfbox{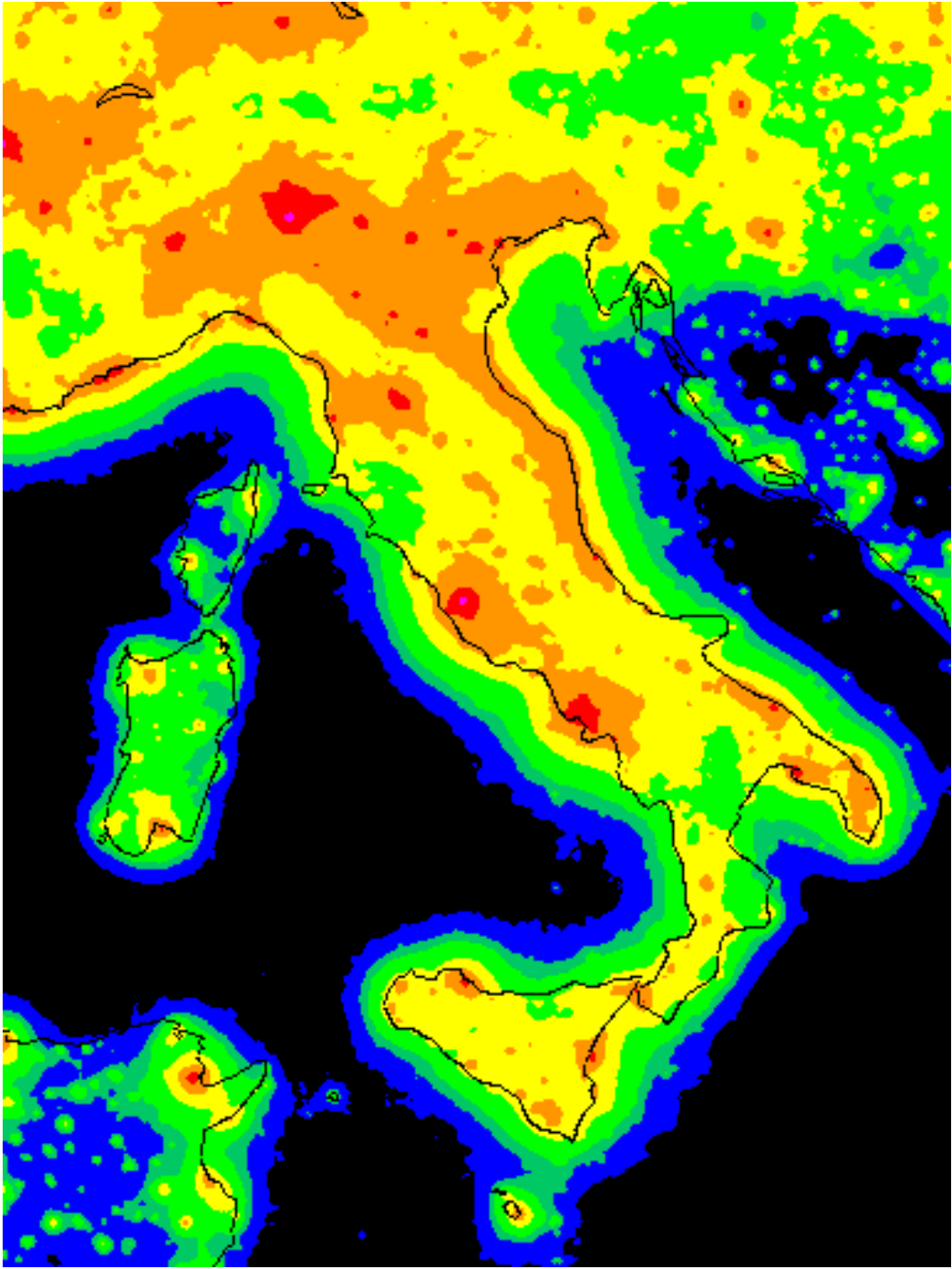} 
\caption[h]{Map of the artificial sky brightness in Italy in 1998. Levels correspond respectively to $<$0.05, 0.05-0.15, 0.15-0.35, 0.35-1.1, 1.1-3, 3-10, 10-30, $>$30 times the natural sky brightness.  Contours of Italy are only an indication.}
\label{fig2}
\end{figure}

\begin{figure}
\epsfysize=19cm 
\hspace{1.5cm}\epsfbox{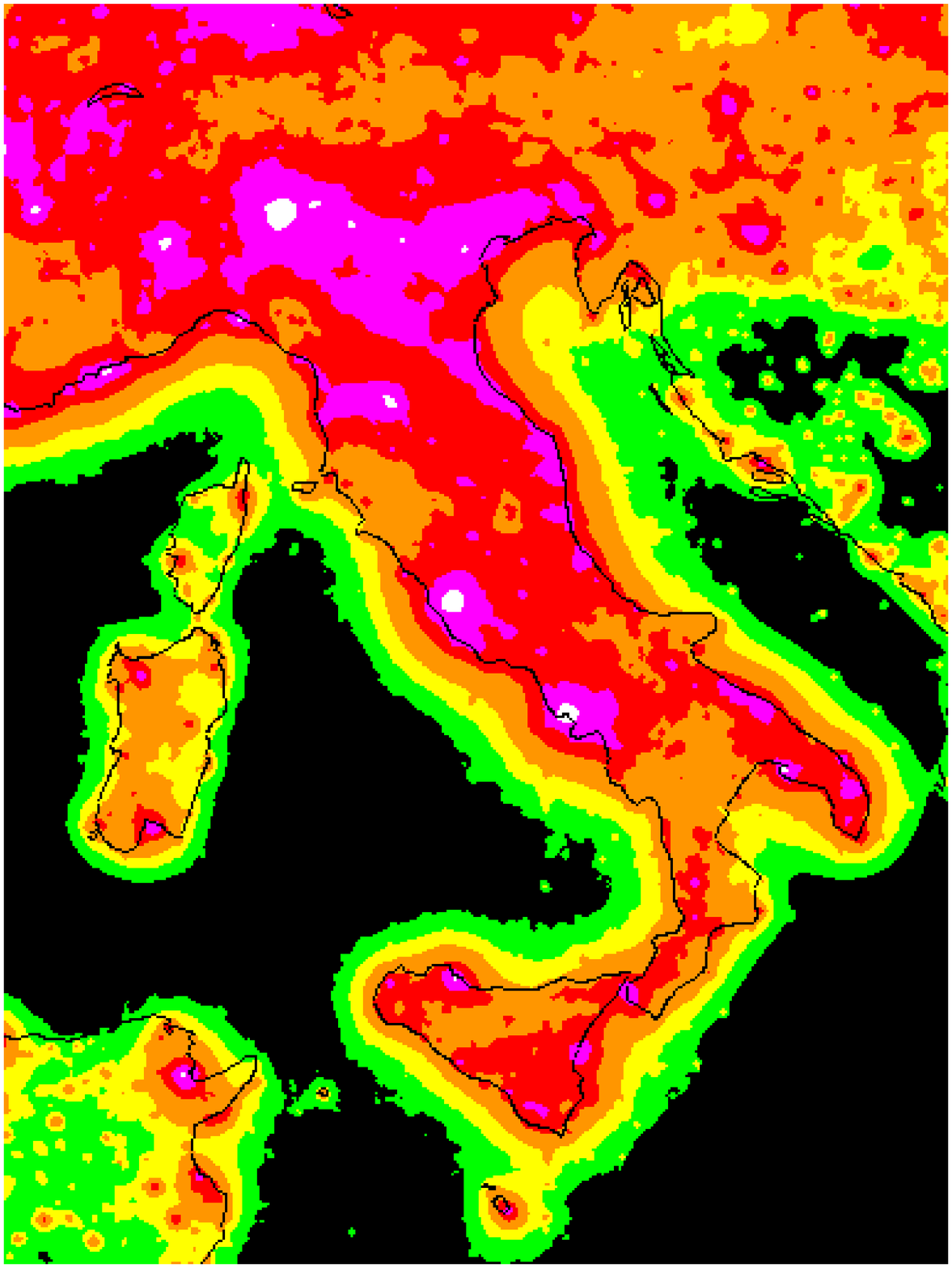} 
\caption[h]{Map of the artificial sky brightness in Italy extrapolated to the year 2025. Levels correspond respectively to $<$0.35, 0.35-1.1, 1.1-3, 3-10, 10-30, 30-100, $>$100 times the natural sky brightness. Contours of Italy are only an indication.}
\label{fig3}
\end{figure}

\begin{equation}  
f((x,y),(x',y'))\propto\left(
\frac{1}{\sqrt{(x-x')^{2}+(y-y')^{2}}}+
\frac{C}{(x-x')^{2}+(y-y')^{2}} \right)
e^{-k \sqrt{(x-x')^{2}+(y-y')^{2}}}  
\end{equation} 
where  C and k are constants related to the relative importance of the direct beam and the scattered component, and the estinction of the city light. The constants were empirically determined by Bertiau et al (1973) and checked by Falchi (1999) fitting the zenith artificial brightness measured at various distances from  some cities in Italy. Differences between B-band and V-band propagation are under the fluctuations given by the atmospheric conditions in standard "clean" nights (Falchi 1999). Details about the input images and the computation have been discussed in Falchi \& Cinzano (1999). In the same paper some maps of artificial and total sky brightness in linear and magnitude scales have been presented. The preliminary calibration was obtained comparing results with available measurements of sky brightness (Falchi \& Cinzano 1999).

In order to compare our results with these of Bertiau \etal (1973) we present in figure \ref{fig1} the map of artificial sky brightness in Italy in 1971 obtained from their data. Levels correspond respectively to these fractions of the natural sky brightness:  $<$0.05, 0.05-0.15, 0.15-0.35, 0.35-1.1, $>$1.1. The resolution of the map is  about 15 km.

In figure \ref{fig2} we present the map of the artificial sky brightness in 1998. Levels are represented as in the previous image but other three levels have been added at 3-10, 10-30, $>$30 times the natural sky brightness in order to show better the situation in the most polluted areas. The true space resolution is about 2.7 km. The comparison of the maps shows as in less than 30 years the artificial sky brightness is increased quite uniformly about 5-10 times, so that now it is greater than the natural one in almost all the italian territory. Only in few small areas of Tuscany, Basilicata, Calabria and Sardinia the natural sky brightness is still greater than the artificial one. No area in Italy has an artificial sky brightness which can be considered acceptable according to IAU recomandations (i.e. lower than 10\% of natural sky brightness - Smith 1979). 
Few areas show an increase over the mean like e.g. the nord-east of Sardinia (Costa Smeralda) where the artificial sky brightness is increased more than 20 times. The average growth rates in Italy are about 6\%-9\% per year in agreement with Cinzano (1999a, 1999b).

Figure \ref{fig3} shows the prediction for the night sky brightness in 2025 extrapolated from the average increase from 1971 to 1998 measured in a sample of 40 sites ($6.3\pm0.3$). We assumed that the trends of light pollution growth remain unchanged. Levels are represented as in the previous images but another level at brightness $>$100 times the natural sky  has been added to better show the brighter areas.  If the average growth rates will remain the same of the last 27 years, we predict that the night sky in 2025 will be 40 times brighter than the sky in 1971.
 
So our conclusion is that the artificial night sky brightness produced by light pollution is a problem which cannot be neglected or delayed anymore. 

\acknowledgements


\end{document}